\documentclass[preprint,12pt]{elsarticle}

\usepackage{rotating}
\usepackage{amssymb}
\usepackage{amsmath}
\usepackage{amsthm}
\usepackage{array}
\usepackage{float}
\usepackage{rotating}
\usepackage{subcaption} 
\usepackage{hyperref}
\begin{document}

\begin{frontmatter}



\title{Urban Dengue and Spatial Dependence: A SAR Model Incorporating Favela Data in Recife, Brazil}


\author{Marcílio Ferreira dos Santos} 

\affiliation{organization={UFPE},
            addressline={}, 
            city={Caruaru},
            postcode={}, 
            state={PE},
            country={Brazil}}

\author{Andreza dos Santos Rodrigues de Melo}
\affiliation{organization={UFPE},
            addressline={}, 
            city={Recife},
            postcode={}, 
            state={PE},
            country={Brazil}}

\begin{abstract}
In this study, we used a dataset comprising approximately 96{,}000 reported dengue cases from 2015 to 2024 in Recife, the capital of northeastern Brazil, all geocoded by neighborhood. The dataset was enriched with sociodemographic data from the national census (population density and average number of residents per household), the proportion of slum areas per neighborhood, a dummy variable identifying neighborhoods with average income above 7.5 minimum wages, the distinction between rainy and dry seasons, and standing/canalized water indicators derived from NDWI and MNDWI indexes. We first tested for spatial dependence in the dependent variable --- dengue cases per square kilometer --- using Moran’s I, which confirmed significant spatial autocorrelation and justified the use of spatial econometric models. Next, we applied the SAR (Spatial Autoregressive) model with these variables and tested four spatial weight structures (Queen, Rook, KNN, and DistanceBand). The best-performing model was SAR with Rook contiguity, which yielded a pseudo-\(R^2\) of 0.6864 and a spatial pseudo-\(R^2\) of 0.6934, indicating strong explanatory power given the multifactorial nature of dengue epidemiology. Five variables showed p-values below 0.05, supporting the statistical significance of their latent effects on disease distribution. After fitting the SAR model, Moran’s I applied to the residuals showed no significant spatial autocorrelation, indicating the model effectively captured the underlying spatial structure. For comparison, we also tested the SAC model using the Rook matrix, but SAR remained slightly superior in performance. In addition to generating a thematic risk map, the SAR model provided a ranking of neighborhoods based on predicted vulnerability. The Spearman correlation between this predicted ranking and the actual case ranking was 0.901, reinforcing the model’s capacity to correctly order neighborhoods by epidemiological risk. These findings confirm the adequacy of the SAR model as a robust analytical tool for understanding spatial patterns of urban arboviruses like dengue, especially when linked to structural socio-environmental conditions such as slum areas, population density, and income. The model enables the creation of dynamic thematic maps and actionable rankings to guide public health efforts. Furthermore, its conclusions can inform stratified SIR models with graph-based representations. 

\end{abstract}

\begin{graphicalabstract}
\includegraphics[width=\textwidth]{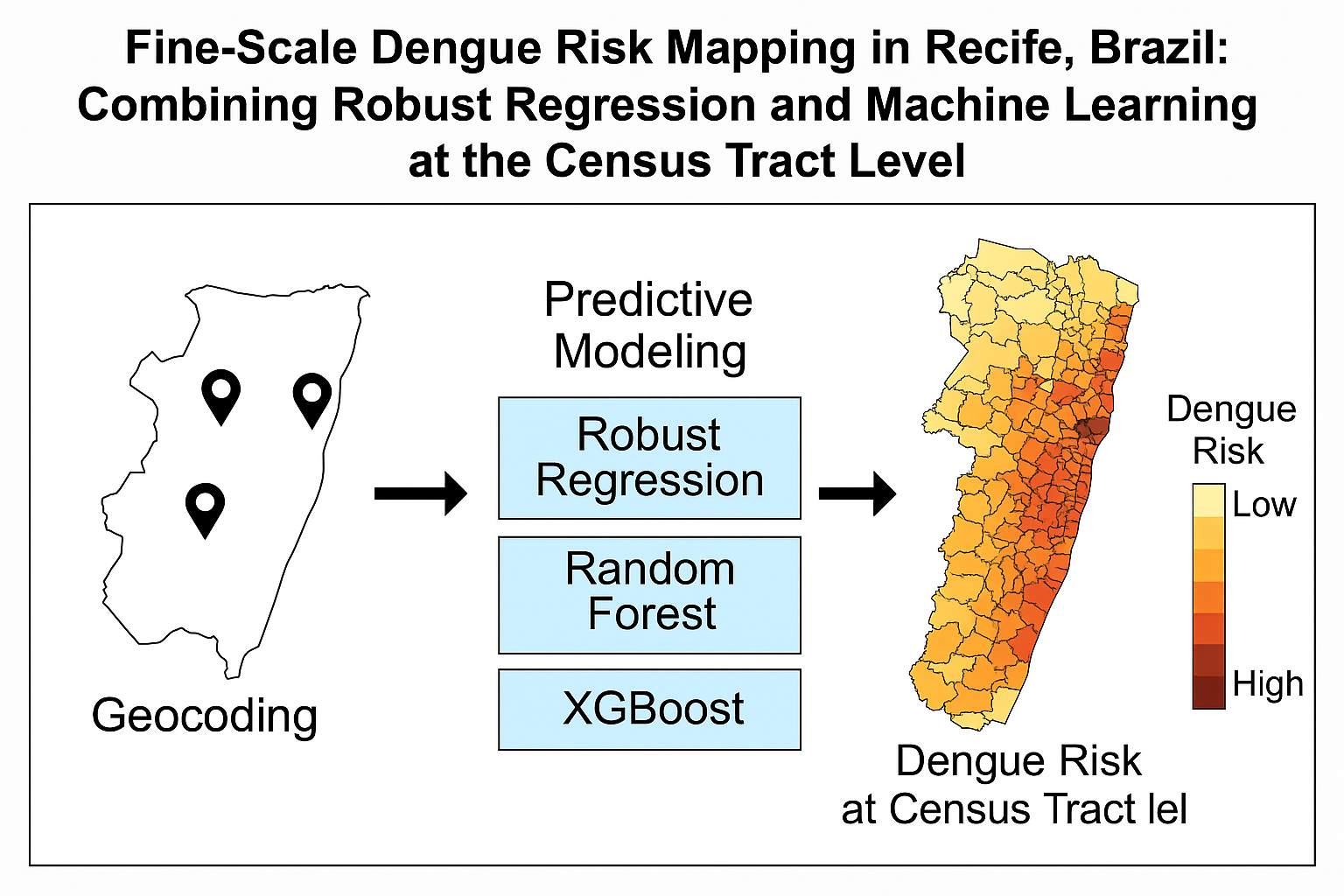}
\end{graphicalabstract}

\begin{highlights}
\item Applied SAR and SAC spatial econometric models to 96,000 dengue cases (2015–2024) in Recife, Brazil.
    \item Models used sociodemographic census data and remote sensing indicators (NDWI/MNDWI).
    \item Rook-based SAR model achieved the best fit (Pseudo-$R^2$ = 0.6864; Spatial Pseudo-$R^2$ = 0.6934).
    \item Residuals showed no spatial autocorrelation (Moran’s I), confirming robustness of the model.
    \item Model-generated rankings strongly correlated with actual data (Spearman $\rho$ = 0.901).
    \item Thematic risk maps and rankings provide guidance for targeted dengue surveillance.
\end{highlights}


\end{frontmatter}



\section{Introduction}

Dengue fever remains one of the major public health challenges in tropical regions, with recurrent outbreaks in urban centers characterized by high population density, precarious housing conditions, and deep socio-spatial inequalities \cite{barcellos2001, andrioli2020}. In Recife—a state capital in Northeast Brazil and one of the most affected cities—these structural factors are exacerbated by unplanned land occupation, deficient urban infrastructure, and a climate highly favorable to the proliferation of \textbf{Aedes aegypti}, with high average temperatures and a rainy season concentrated between March and July \cite{ferreira2022, borges2024}. Additionally, climatic variations such as El Niño have shown a significant impact on the intensification of cases \cite{ferreira2022}.

National studies highlight that Brazil leads the world in reported dengue cases, and identify Recife as a persistent hotspot since the reemergence of the disease in the 1980s, due to a combination of high population density, unregulated urban growth, and insufficient sanitation services \cite{teixeira2009}. The city exhibits a substantial presence of informal settlements, an inadequate urban drainage system, and marked hydrological seasonality—elements that contribute to vector persistence and hinder effective disease control. This scenario underscores the need for refined territorial approaches aimed at identifying critical areas and guiding targeted public health interventions.

Although Brazil offers a wide range of epidemiological and census data, intra-urban studies—particularly at the neighborhood level—remain scarce, especially those employing statistical models that explicitly incorporate spatial dependence between geographic units \cite{perles2021, honorio2009}. This methodological gap limits the detection of local risk patterns and constrains the design of spatially targeted surveillance and control strategies.

Spatial econometric models, such as the Spatial Autoregressive Model (SAR), are well suited to capture spatial autocorrelation and spillover effects between neighboring regions \cite{anselin1988, anselin1995, lesage2009, souza2001}. These models are especially relevant in urban settings, where the spread of arboviruses is closely linked to human mobility, vector ecology, and shared socio-environmental characteristics.

In this study, we applied both SAR and SAC (Spatial Autoregressive Combined) models to account for different forms of spatial dependence in the distribution of dengue cases in Recife. While the SAR model incorporates spatial autocorrelation in the dependent variable, the SAC model simultaneously accounts for correlation in both the dependent variable and the residuals \cite{elhorst2014}.

Our modeling was based on approximately 96,000 georeferenced confirmed dengue cases recorded between 2015 and 2024, provided by the Recife Municipal Informatics Agency (EMPREL)\footnote{\url{https://www.emprel.gov.br/}}. The dataset was enriched with variables from the 2022 Demographic Census (six variables), a neighborhood-level metric of informal settlements, and environmental and climatic indicators derived from remote sensing. To reduce multicollinearity, we applied Principal Component Analysis (PCA) \cite{baldoquin2023}, synthesizing ten explanatory variables into statistically independent components (including additional census data such as vacant and collective dwellings). However, PCA did not improve regression performance compared to direct use of the original variables, which were highly interpretable and had strong explanatory power.

The results demonstrate that the SAR model using a Rook spatial weights matrix achieved the best performance, with a pseudo $R^2$ of up to 68.64\%. The more complex SAC model did not yield significant improvements over SAR, indicating that the spatial structure of the dependent variable alone sufficiently explained the observed patterns. Furthermore, residuals showed no significant spatial autocorrelation (Moran’s I), confirming the statistical adequacy of the models.

The main contribution of this study lies in the construction of a robust and explainable spatial model, supported by a comprehensive empirical dataset and a replicable methodological approach. The inclusion of less common variables, such as the proportion of informal settlements and NDWI/MNDWI indices \cite{mudele2021modeling}, along with neighborhood-level analysis, enabled accurate dengue risk mapping and territorial ranking, with a high correlation (0.901) between predicted values and actual case data. The findings provide valuable empirical support for evidence-based public policy planning and offer a foundation for future approaches integrating complex network analysis and spatially structured SIR compartmental models.

\section{Spatial Autoregressive Model (SAR)}

The Spatial Autoregressive Model (SAR) is one of the main tools in spatial econometrics for explicitly capturing spatial dependence between geographic units \cite{anselin1988, lesage2009, elhorst2014}. In urban epidemiological contexts—such as dengue outbreaks—this modeling approach is particularly useful for representing spatial diffusion patterns, feedback between adjacent neighborhoods, and territorial spillover effects.

The general formulation of the SAR model is given by:

\begin{equation}
    \mathbf{y} = \rho \mathbf{W}\mathbf{y} + \mathbf{X}\boldsymbol{\beta} + \boldsymbol{\varepsilon}
\end{equation}

where:
\begin{itemize}
    \item $\mathbf{y}$ is the vector of observations of the dependent variable (in this study, dengue case density by neighborhood);
    \item $\mathbf{X}$ is the matrix of explanatory variables;
    \item $\boldsymbol{\beta}$ is the vector of regression coefficients;
    \item $\rho$ is the spatial autocorrelation coefficient of the dependent variable;
    \item $\mathbf{W}$ is the row-normalized spatial weight matrix;
    \item $\boldsymbol{\varepsilon}$ is the vector of random error terms.
\end{itemize}

The parameter $\rho$ plays a central role in model interpretation: when statistically significant, it indicates that the values of the dependent variable in a given territorial unit are systematically related to the values observed in neighboring units. This spatial pattern may reflect mechanisms of direct contagion as well as shared environmental, demographic, or social conditions—all of which are relevant to the spread of urban arboviruses.

Defining the spatial weight matrix $\mathbf{W}$ is a critical step in the modeling process, as it determines the neighborhood structure and directly affects the estimated parameters. In this study, we tested four classical spatial configurations:

\begin{enumerate}
    \item \textbf{Queen contiguity}: considers as neighbors those units that share at least one point (edge or vertex);
    \item \textbf{Rook contiguity}: considers as neighbors only those units that share a common boundary;
    \item \textbf{K-nearest neighbors (KNN)}: connects each unit to its $k$ closest units based on Euclidean distance between centroids;
    \item \textbf{Fixed distance}: defines as neighbors all units located within a fixed geographic radius.
\end{enumerate}

Comparing these different neighborhood structures allows for assessing the model’s sensitivity to the adopted spatial configuration—an especially relevant aspect in urban settings like Recife, where patterns of population density, urban infrastructure, and social vulnerability vary significantly across neighborhoods. The city's socioeconomic morphology is marked by \textit{“islands”} of high income surrounded by low-income areas, which intensify daily commuting flows of workers and services between socially contrasting zones \cite{massaro2019}. These mobility patterns shape a functional connectivity that is not always captured by purely geographic criteria, reinforcing the importance of incorporating real territorial dynamics into the construction of spatial weight matrices. This approach seeks to produce a more sensitive and context-aware model, aligned with the complexity of dengue transmission in socially unequal urban environments.

\begin{figure}[H]
    \centering
    \includegraphics[width=1\textwidth]{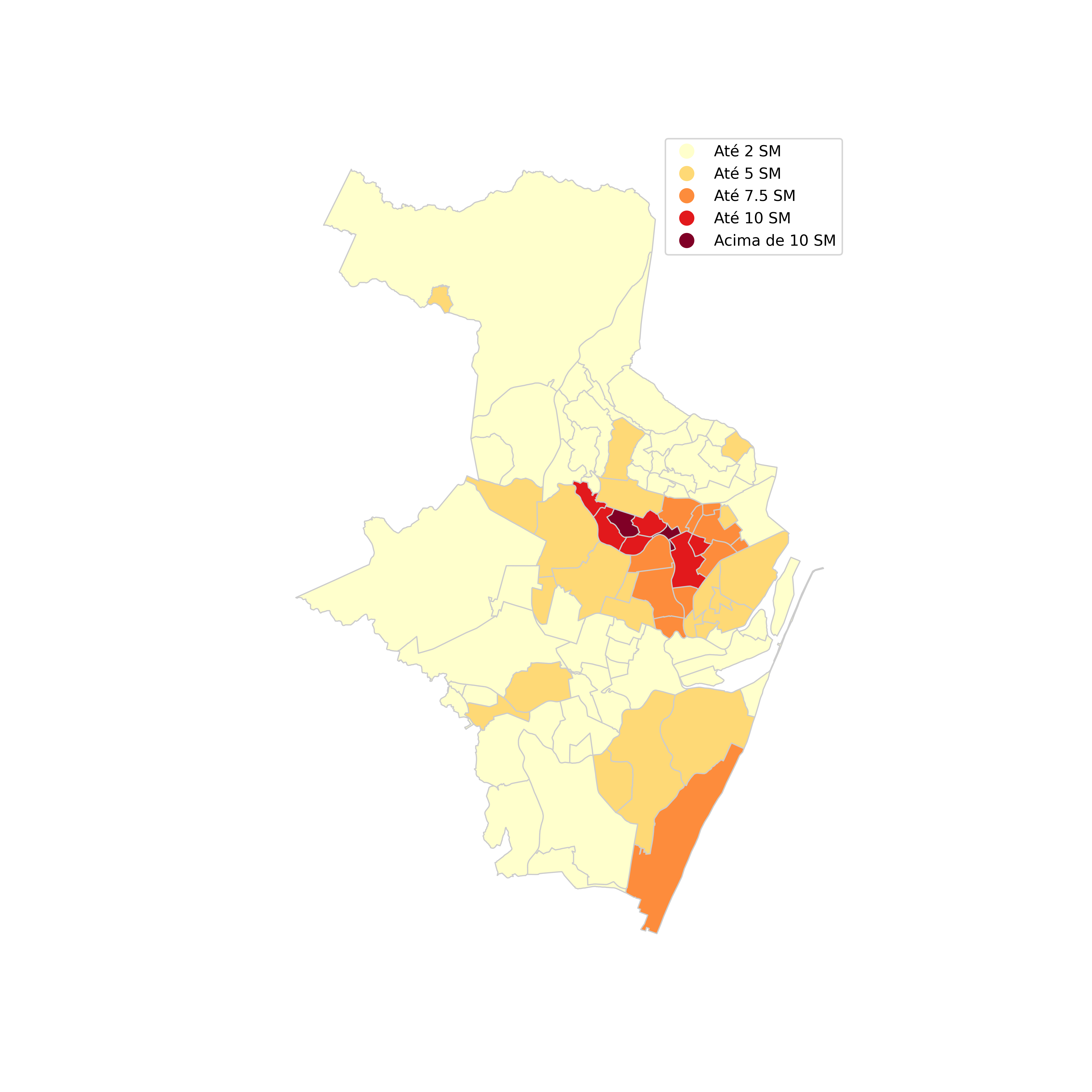}
    \caption{Average household income by neighborhood in Recife (2022 Census). Source: IBGE / Author's elaboration.}
    \label{fig:map_income}
\end{figure}

To illustrate Recife's heterogeneous socioeconomic morphology, Figure~\ref{fig:map_income} presents a map of average household income distribution by neighborhood, based on data from the 2022 Demographic Census. One can observe the presence of high-income "islands"—particularly in the southern and central areas of the city—surrounded by neighborhoods with substantially lower economic indicators. This configuration reinforces the need for models that not only account for geographic proximity, but also capture territorial dynamics associated with daily commuting and functional connectivity across socially contrasting zones.

\section{Spatial Autoregressive Combined Model (SAC)}

The Spatial Autoregressive Combined model (SAC) extends the SAR and SEM specifications by simultaneously incorporating spatial autocorrelation in both the dependent variable and the residuals \cite{lesage2009, elhorst2014}. This structure is particularly useful when there is evidence of multiple spatial processes acting concurrently—such as direct interdependence between territorial units and unobserved latent effects.

The general formulation of the SAC model is given by:

\begin{equation}
    \mathbf{y} = \rho \mathbf{W}\mathbf{y} + \mathbf{X}\boldsymbol{\beta} + \mathbf{u}, \quad \text{with} \quad \mathbf{u} = \lambda \mathbf{W}\mathbf{u} + \boldsymbol{\varepsilon}
\end{equation}

where:
\begin{itemize}
    \item $\rho$ captures spatial autocorrelation in the dependent variable (SAR component);
    \item $\lambda$ represents spatial autocorrelation in the residuals (SEM component);
    \item $\mathbf{W}$ is the spatial weights matrix;
    \item $\boldsymbol{\varepsilon}$ is a vector of i.i.d. error terms.
\end{itemize}

The presence of two distinct spatial parameters ($\rho$ and $\lambda$) provides the SAC model with greater flexibility to describe complex spatial dependence patterns. This is particularly relevant in heterogeneous urban settings such as Recife, where dengue transmission may result from both structural conditions (e.g., urban density, sanitation) and latent externalities not captured by the explanatory variables.

In this study, we apply the SAC model using the spatial weights matrix that yielded the best performance in the SAR model. This strategy allows us to test whether modeling residual spatial structure adds explanatory power, while avoiding model overfitting and preserving comparability across specifications—an approach aligned with methodological best practices in spatial econometrics \cite{elhorst2014}.

\section{Model Fit Assessment: Traditional and Spatial Pseudo-\texorpdfstring{$R^2$}{R²}}

Evaluating the goodness-of-fit of spatial models — such as the SAR and SAC models — requires metrics that consider not only the overall explanatory power but also the model’s ability to capture spatial dependence structures. To this end, two complementary measures are widely used: the traditional \textit{Pseudo-}$R^2$ and the spatial \textit{Pseudo-}$R^2$.

As emphasized by LeSage and Pace \cite{lesage2009} and Elhorst \cite{elhorst2014}, relying solely on classical linear regression metrics can be inadequate in spatial contexts because they ignore interactions between neighboring geographic units. These interactions, known as spatial spillovers, reflect the spillover effects across regions: the value of a variable in one territorial unit may be influenced by characteristics or events occurring in adjacent units. In the case of vector-borne diseases such as dengue, these effects may relate to urban mobility, environmental continuity, or human circulation between connected neighborhoods.

Below we detail the two metrics employed in this study.

\subsection{Traditional Pseudo-\texorpdfstring{$R^2$}{R²}}

The traditional \textit{pseudo-}$R^2$ is an extension of the classical coefficient of determination, adapted for models in which residual structures may violate the assumptions of ordinary linear regression. It measures the proportion of variability in the dependent variable $y$ explained by the fitted model, without explicitly accounting for spatial autocorrelation. Its basic formula is:

\begin{equation}
\text{pseudo-}R^2 = 1 - \frac{\sum_{i} (y_i - \hat{y}_i)^2}{\sum_{i} (y_i - \bar{y})^2}
\end{equation}

While useful for assessing overall explanatory power, this metric can be insufficient in spatial settings because it ignores neighborhood structure and the spatial dependence effects inherent in the data dynamics.

\subsection{Spatial Pseudo-\texorpdfstring{$R^2$}{R²}}

The spatial \textit{Pseudo-}$R^2$ aims to quantify how much of the spatial structure present in the data is captured by the model. It explicitly accounts for direct and indirect spillover effects — that is, the influences that neighboring units exert on a specific unit, as well as the influences those neighbors receive from other units.

Two common formulations include:

\begin{equation}
R^2_{\text{spatial}} = 1 - \frac{\mathrm{var}(\hat{u})}{\mathrm{var}(y)}
\end{equation}

\begin{equation}
R^2_{\text{spatial}} = \left( \mathrm{corr}(y, \hat{y}_{\text{spatial}}) \right)^2
\end{equation}

These measures evaluate the model’s ability to represent the underlying spatial dependence in the data, which is crucial in applications such as urban epidemiology, health geography, territorial planning, and evidence-based public policy.

\subsection{Comparison and Recommendations}

Table~\ref{tab:comparacao_r2} summarizes the key differences between the two indicators. Practically, it is recommended to use both the traditional and spatial \textit{Pseudo-}$R^2$ metrics in combination: the former assesses the proportion of the phenomenon explained globally; the latter reveals whether the observed spatial patterns are well captured by the model.

\begin{sidewaystable}
\centering
\caption{Comparison between Traditional and Spatial Pseudo-$R^2$}
\label{tab:comparacao_r2}
\begin{tabular}{|p{4.5cm}|p{6cm}|p{6cm}|}
\hline
\textbf{Criterion} & \textbf{Traditional Pseudo-$R^2$} & \textbf{Spatial Pseudo-$R^2$} \\
\hline
Explained variance & Measures the proportion of total variance of the dependent variable explained by the model. & Measures the proportion of spatial structure in the data captured by the model. \\
\hline
Accounts for spatial autocorrelation? & Does not consider relationships among neighboring units. & Explicitly incorporates spatial dependence. \\
\hline
Applicable models & Usable in OLS and SAR/SAC models. \newline However, limited in spatial models. & Recommended only for models with spatial structure (SAR, SAC, etc.). \\
\hline
Similarity to classic $R^2$ & Direct interpretation similar to classical linear regression $R^2$. & Partial interpretation, as it involves both direct and indirect spatial effects. \\
\hline
Recommended use & To evaluate the overall goodness-of-fit of the model. & To verify whether spatial patterns are well represented by the model. \\
\hline
\end{tabular}
\end{sidewaystable}

In this study, both metrics served as central criteria for evaluating and comparing the SAR and SAC models, testing different spatial weight matrices (queen, rook, KNN, and fixed distance). The results guided the selection of the most appropriate model, balancing statistical explanatory power with sensitivity to the spatial structure of dengue dissemination in Recife.

\section{Study Design and Methodology}

\subsection{Study Area and Data Sources}

This study focuses on the city of Recife, the capital of the state of Pernambuco, located on the northeastern coast of Brazil. With an estimated population of approximately 1.48 million inhabitants and an urban area of around 218 km², the municipality is marked by high population density, pronounced socioeconomic heterogeneity, and environmental vulnerabilities that facilitate the transmission of arboviruses such as dengue \cite{ibge2024, rio2007}.

The dataset comprises confirmed dengue cases from 2015 to 2024, made available through the open data platform of Recife’s municipal government (EMPREL). These records include information on the neighborhood of residence, year of diagnosis, sex, clinical symptoms, and educational level of the patients. Geocoding was conducted based on postal codes and street addresses using the ViaCEP API. The data were then spatially aggregated at the neighborhood centroid level to ensure anonymity, in accordance with Brazil’s General Data Protection Law (LGPD).

Socioeconomic variables were extracted from the 2022 Demographic Census conducted by the Brazilian Institute of Geography and Statistics (IBGE), encompassing indicators related to population density, household composition, urban verticalization, sanitary conditions, and access to infrastructure. A detailed description of these variables is provided in Appendix A.

In addition, three environmental variables were incorporated to capture territorial conditions potentially associated with the presence of the Aedes aegypti mosquito: vegetation cover, the presence of open canals, and informal urban settlements (favelas). Each variable was expressed as the proportion of the neighborhood's area covered by these features. This standardization allows for consistent comparisons across neighborhoods of different sizes.

\begin{figure}[H]
\centering
\includegraphics[width=0.5\linewidth]{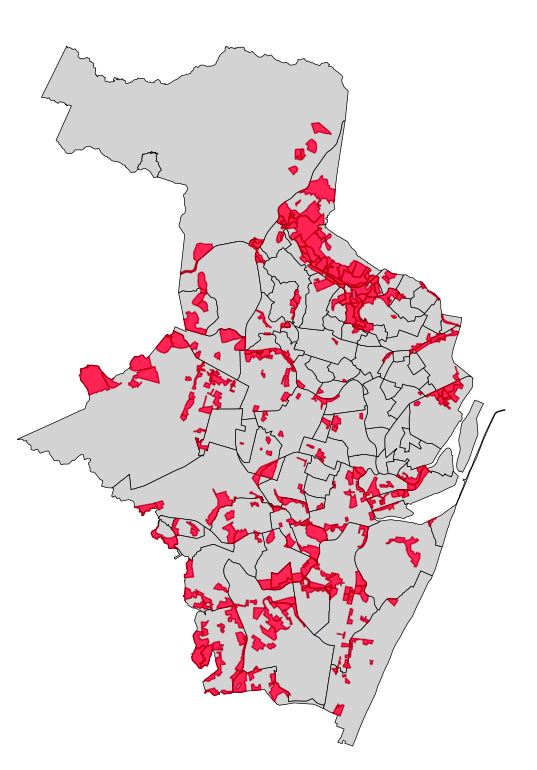}
\caption{Distribution of informal settlement areas (favelas) across neighborhoods in Recife.\\
\small Source: elaborated by the author based on data from IBGE \cite{ibge2023asn}.}
\label{fig:favelas}
\end{figure}

\begin{figure}[htbp]
\centering
\begin{subfigure}[t]{0.32\textwidth}
\includegraphics[width=\linewidth]{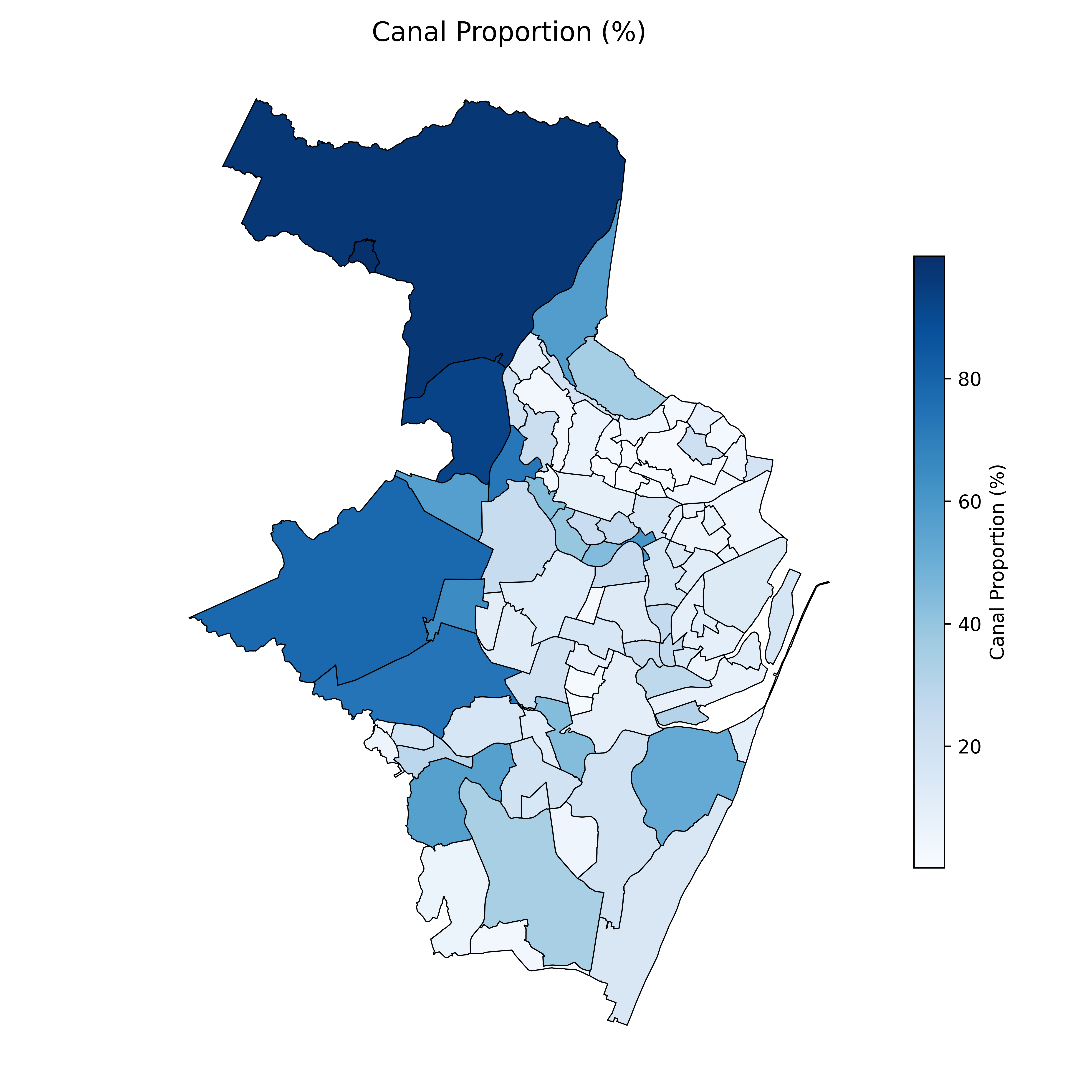}
\caption{Canal proportion}
\label{fig:map_canal}
\end{subfigure}
\hfill
\begin{subfigure}[t]{0.32\textwidth}
\includegraphics[width=\linewidth]{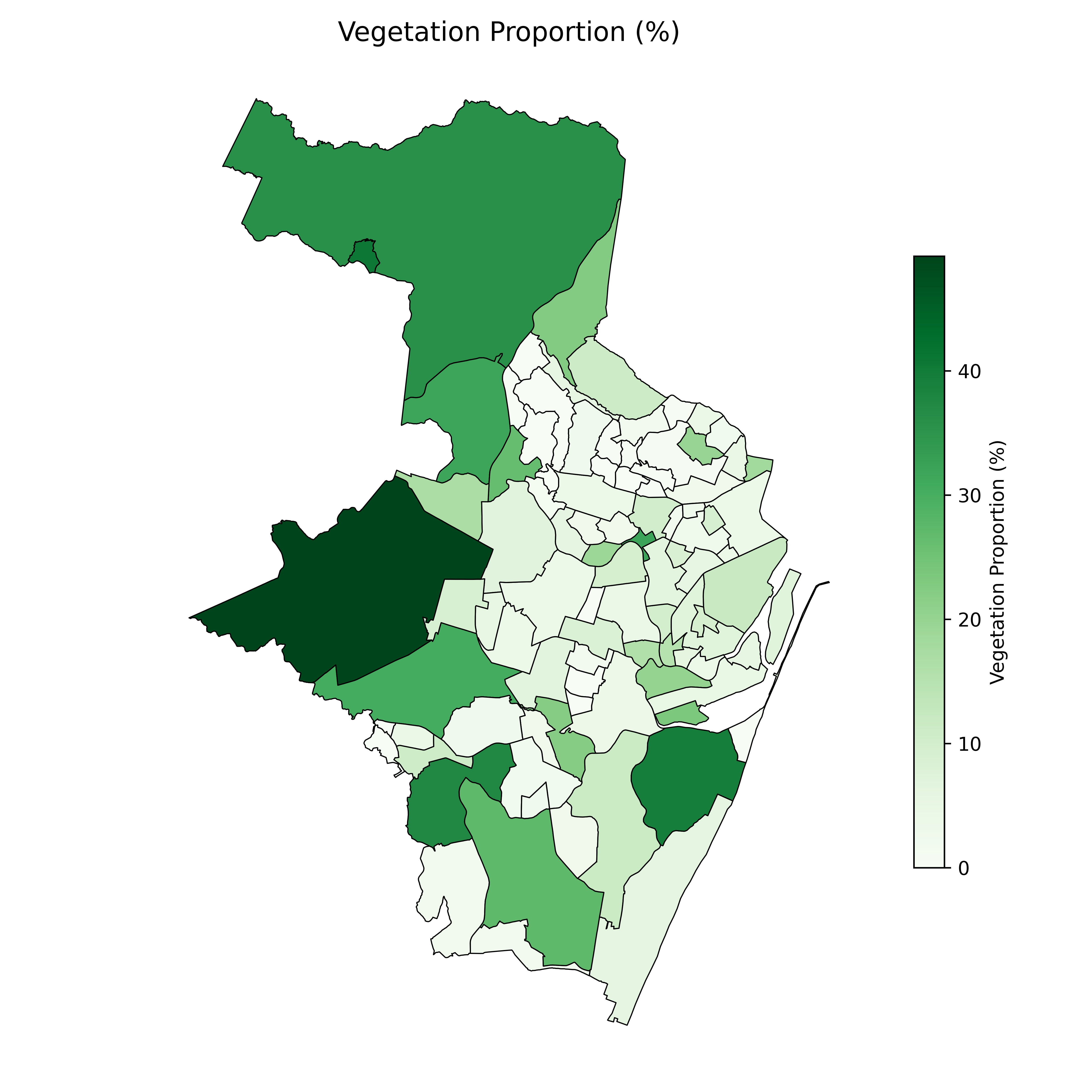}
\caption{Vegetation proportion}
\label{fig:mapa_vegetacao}
\end{subfigure}
\hfill
\begin{subfigure}[t]{0.32\textwidth}
\includegraphics[width=\linewidth]{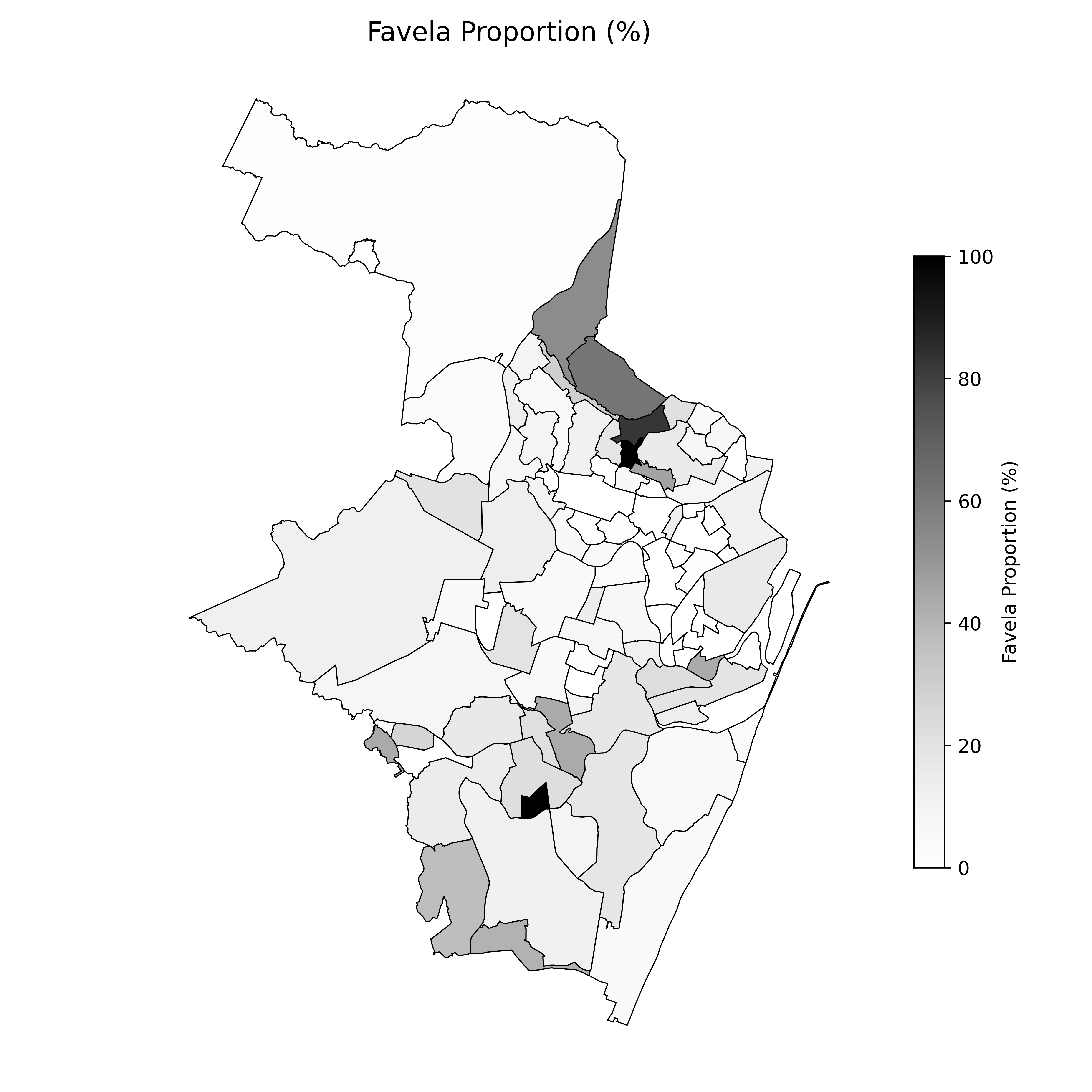}
\caption{Favela proportion}
\label{fig:mapa_favela}
\end{subfigure}
\caption{Thematic maps showing the proportion of canals, vegetation, and informal settlements (favelas) in Recife neighborhoods.}
\label{fig:thematic_maps}
\end{figure}

Figures~\ref{fig:thematic_maps} present thematic maps depicting the spatial distribution of canals, vegetation, and informal settlements by neighborhood. The canal variable represents the presence of natural or artificial water bodies that may influence both water runoff and retention, directly impacting mosquito breeding dynamics.

Vegetation cover was estimated using satellite imagery and validated using the Normalized Difference Vegetation Index (NDVI). However, a derived variable — calculated as the difference between the Normalized Difference Water Index (NDWI) and NDVI, known as the Modified NDWI (MNDWI) — showed superior performance in predicting dengue incidence. This composite index is strongly correlated with shallow water bodies, puddles, and water-accumulation zones, which serve as optimal breeding habitats for mosquitoes.

Interestingly, the MNDWI-derived variable was negatively associated with dengue incidence—that is, neighborhoods with higher vegetation and moisture signals tended to report fewer cases. This relationship may reflect the protective role of environmental features such as dense vegetation, higher humidity, and stable urban microclimates, which can hinder the Aedes aegypti life cycle. Prior studies have indicated that high temperatures accelerate mosquito development and egg hatching, whereas vegetated areas may function as thermal and physical barriers \cite{simoy2015, adeola2017, parselia2019}.

Thus, the use of MNDWI as an environmental covariate proved promising in this study, capturing both potential sources of stagnant water and ecological attributes that mediate disease transmission risk—enhancing the interpretation of environmental determinants in the spread of dengue.

\subsection{Analytical Procedures}

To reduce multicollinearity among census-derived indicators, Principal Component Analysis (PCA) was initially conducted as an exploratory step. However, the final models were estimated using the original variables, as PCA did not yield a meaningful improvement in predictive accuracy.

The spatial modeling approach adopted in this study exclusively relied on spatial regression techniques to account for geographic dependencies in dengue distribution. The primary model employed was the Spatial Autoregressive Model (SAR), which incorporates a spatial lag of the dependent variable, enabling estimation of spillover effects between adjacent neighborhoods.

Parameter estimation was carried out via maximum likelihood for the SAR model and using the GM-2SLS method for the Spatial Autoregressive Combined (SAC) model, with implementation through the \texttt{spreg} and \texttt{libpysal} Python packages. Four spatial weight matrix specifications were tested: Queen contiguity, Rook contiguity, first-order \textit{k}-nearest neighbors, and fixed distance. Among these, the Queen and Rook matrices outperformed the others, with Rook showing a slight advantage.

To assess spatial dependence before and after model fitting, Moran's I statistic was applied both to the dependent variable and to the residuals. This allowed for verification of whether the spatial autocorrelation was adequately captured by the models. The SAC model was also tested to evaluate potential improvements in explanatory power and residual independence.

The explanatory variables used in the final models are described in Table~\ref{tab:variables}.

\begin{table}
\centering
\caption{Description of explanatory variables used in the model}
\label{tab:variables}
\begin{tabular}{@{}lp{7.5cm}@{}}
\hline
\textbf{Variable} & \textbf{Description} \\
\hline
Population Density        & Number of inhabitants per square kilometer in each neighborhood. \\
Rainy Season Indicator   & Binary indicator for months falling within the local rainy season. \\
Average Household Size   & Average number of residents per household. \\
Canal Proportion          & Proportion of neighborhood area covered by open drainage canals. \\
Favela Proportion         & Proportion of neighborhood area occupied by informal settlements. \\
Higher Income Proportion  & Proportion of residents classified in the highest income bracket. \\
\hline
\end{tabular}
\end{table}

\subsection{Risk Ranking and Empirical Analysis}

Based on predictions generated by the spatial models, a neighborhood-level dengue risk ranking was constructed using the estimated density of cases. This modeled ranking was then compared to the empirical distribution of cases reported in 2024 to validate its predictive performance using only census and environmental data.

Furthermore, a detailed examination of the socio-spatial profiles of neighborhoods with the highest and lowest modeled risk was conducted to provide empirical grounding for the model outputs and generate policy-relevant territorial insights.

Using these results, thematic maps were created to visually represent modeled risk concentration levels. These maps are proposed as decision-support tools for targeting public health interventions and guiding spatially informed policy actions.

\section{Resultados}

\subsection{Spatial Autocorrelation Analysis}
Prior to implementing spatial regression models, a global spatial autocorrelation test — Moran's I — was conducted for the dependent variable, defined as the number of dengue cases per square kilometer in each neighborhood of Recife. The results, presented in Table~\ref{tab:moran_results}, indicated a positive and statistically significant spatial autocorrelation. This finding confirms the presence of spatial clustering of cases and supports the application of models that incorporate spatial dependence in their parameter structures.

\begin{table}[H]
\centering
\caption{Global spatial autocorrelation test (Moran’s I) for the variable \textit{casos\_p\_km2}, using different spatial weight matrices.}
\label{tab:moran_results}
\begin{tabular}{>{\raggedright\arraybackslash}p{6.5cm}cc}
\hline
\textbf{Spatial Weight Matrix} & \textbf{Moran’s I} & \textbf{p-value} \\
\hline
Queen Contiguity (vertices and edges) & $0.2699$ & $0.0010$ \\
Rook Contiguity (edges only) & $0.2789$ & $0.0010$ \\
K-Nearest Neighbors ($k = 5$, UTM projection) & $0.2762$ & $0.0010$ \\
Distance Band (radius = 1000 m, UTM projection) & $0.1999$ & $0.0030$ \\
\hline
\end{tabular}
\end{table}

\noindent\textit{Note on spatial projections:} The contiguity-based matrices (Queen and Rook) were constructed using geographic data in the EPSG:4674 system (SIRGAS 2000), suitable for geodetic coordinates. In contrast, distance-based matrices (KNN and Distance Band) require a metric projection — such as the UTM system — to ensure accurate distance measurements. For these analyses, the data were properly reprojected, ensuring the validity of the spatial relationships.

\subsection{Detection of Stagnant Water Using Spectral Difference}

As part of the incorporation of environmental variables, a spectral approach was employed to identify urban water bodies with a higher potential for stagnation, such as canals, puddles, and flood-prone areas.

The methodology is based on the difference between two widely used spectral indices:

\begin{itemize}
\item \textbf{NDWI} (Normalized Difference Water Index): uses the green (B3) and near-infrared (B8) bands, being sensitive to the presence of water but also affected by shadows, built surfaces, and moist soils.
\item \textbf{MNDWI} (Modified NDWI): uses the green (B3) and shortwave infrared (B11) bands, which are more effective in detecting water bodies in urban contexts, with reduced spectral interference from built-up areas.
\end{itemize}

By computing the difference \textbf{MNDWI $-$ NDWI}, the method enhances the spectral response of water features that are strongly detected by MNDWI but weakly by NDWI — a characteristic commonly associated with low-flow water bodies, greater turbidity, or presence within vegetated regions. A threshold (> 0.05) was applied to segment areas likely to contain urban canals or stagnant water features.

This spectral difference proved to be an effective predictor of potentially flooded areas and was significantly correlated with the epidemiological variable of interest. It was therefore incorporated into the spatial regression as a proxy for the presence of non-flowing aquatic breeding sites.

\subsection{Comparison of Neighborhood Matrices in the SAR Model}

This section presents the results of applying SAR (Spatial Autoregressive) models using different spatial weight matrices: Queen, Rook, K-Nearest Neighbors (KNN), and Distance Band. The goal was to compare model performance in terms of explanatory power and stability of coefficient estimates for the dependent variable \textit{units\_p\_km2}.

The variable \textit{units\_p\_km2} represents the spatial density of dengue cases, calculated as the number of confirmed cases per square kilometer in each neighborhood. This standardization corrects for distortions caused by differences in area between spatial units, allowing a fairer comparison of disease intensity across the territory.

\begin{table}[H]
\centering
\caption{Desempenho dos modelos SAR com diferentes matrizes de pesos espaciais.}
\label{tab:sar_comparison}
\renewcommand{\arraystretch}{1.2} 
\begin{tabular}{>{\raggedright\arraybackslash}p{2.8cm} 
                >{\centering\arraybackslash}p{2cm} 
                >{\centering\arraybackslash}p{3.2cm} 
                >{\centering\arraybackslash}p{1.8cm} 
                >{\centering\arraybackslash}p{1.8cm}}
\hline
\textbf{Matriz de Pesos} & \textbf{Pseudo $R^2$} & \textbf{Log-Verossimilhança} & \textbf{AIC} & \textbf{Coef. $W y$} \\
\hline
Queen        & 0{,}6859 & -1222{,}53 & 2461{,}06 & 0{,}1392 \\
Rook              & \textbf{0{,}6864} & \textbf{-1222{,}41} & \textbf{2460{,}82} & \textbf{0{,}1468} \\
KNN ($k$ = 5)               & 0{,}6840 & -1222{,}99 & 2461{,}99 & 0{,}0847 \\
Banda de Distância                & 0{,}6821 & -1223{,}48 & 2462{,}96 & -0{,}0273 \\
\hline
\end{tabular}
\end{table}

Models using contiguity matrices (Queen and Rook) exhibited better performance in terms of Pseudo $R^2$, log-likelihood, and AIC. The Rook matrix yielded slightly superior results, with better overall fit and greater statistical significance for the spatial lag coefficient ($W y$), indicating stronger spatial dependence captured.

Conversely, models based on distance (KNN and Distance Band) showed weaker performance. In the case of the Distance Band, the spatial coefficient was negative and non-significant, suggesting that this matrix does not adequately capture the spatial relationships relevant to the variable of interest.

Table~\ref{tab:sar_coefficients} presents the estimated coefficients of explanatory variables across the SAR models.

\begin{table}[H]
\centering
\caption{Coeficientes estimados nos modelos SAR com diferentes matrizes espaciais.}
\label{tab:sar_coefficients}
\renewcommand{\arraystretch}{1.2} 
\begin{tabular}{
  >{\raggedright\arraybackslash}p{3cm}  
  >{\centering\arraybackslash}p{2cm}     
  >{\centering\arraybackslash}p{2cm}     
  >{\centering\arraybackslash}p{2cm}     
  >{\centering\arraybackslash}p{2cm}     
}
\hline
\textbf{Variável} & \textbf{Queen} & \textbf{Rook} & \textbf{KNN} & \textbf{Banda de Distância} \\
\hline
Constante & -770{,}06 & -762{,}79 & -787{,}95 & -836{,}83 \\
Densidade Populacional & 0{,}0231 & 0{,}0231 & 0{,}0231 & 0{,}0241 \\
Período Chuvoso & 222{,}55 & 222{,}80 & 223{,}53 & 220{,}05 \\
Média Moradores por Domicílio & 320{,}06 & 316{,}14 & 336{,}45 & 368{,}90 \\
Proporção de Canais & -368{,}62 & -365{,}57 & -392{,}33 & -417{,}98 \\
Proporção de Favelas & -224{,}41 & -223{,}11 & -233{,}27 & -226{,}02 \\
Mais Rico & -189{,}78 & -188{,}53 & -193{,}78 & -203{,}24 \\
Coef. $W y$ & 0{,}1392 & 0{,}1468 & 0{,}0847 & -0{,}0273 \\
\hline
\end{tabular}
\end{table}

All variable coefficients retained consistent signs across models, reinforcing the robustness of the findings. Population density, rainy season, and the mean number of residents per household showed positive associations with dengue incidence, while the proportions of canals, slums, and high-income population showed negative associations.

The negative coefficient for the \textit{proportion of canals} variable, though initially counterintuitive, may reflect spatial overlap with densely vegetated areas — such as mangroves — in neighborhoods like Várzea and Guabiraba. A high correlation between vegetation and canals ($r = 0.85$) supports this hypothesis.

These findings align with studies highlighting the moderating role of vegetation on dengue risk, as it provides higher humidity, shading, and microclimates less favorable to mosquito reproduction \cite{zellweger2017, abdullah2025}. Thus, even regions containing water bodies may exhibit protective effects when associated with dense vegetation cover.

These results confirm the suitability of the SAR model for spatial analysis of dengue in Recife, with the Rook matrix offering the best performance. Matrix selection should consider both statistical fit and the territorial coherence of spatial relationships among neighborhoods.

\subsection{Residual Analysis: Remaining Spatial Autocorrelation}
Following the calibration of the SAR (Spatial Autoregressive) models with different spatial weight matrices (Queen, Rook, K-Nearest Neighbors, and Distance Band), we conducted an analysis of the spatial autocorrelation of the residuals. The goal was to assess whether the spatial dependence present in the data had been adequately accounted for.

To this end, the Moran’s I statistic was computed for the residuals of each model. This index ranges from -1 to 1: values near zero indicate spatial randomness (i.e., absence of spatial autocorrelation), positive values suggest the presence of spatial clustering, and negative values imply spatial dispersion. Statistical significance was assessed using permutation tests.

The results are presented in Table~\ref{tab:moran_residuos}. All models yielded residuals with Moran’s I values close to zero and non-significant p-values, indicating the absence of remaining spatial autocorrelation. This supports the adequacy of the fitted SAR models and strengthens the statistical validity of the inference process.

\begin{table}[H]
\centering
\caption{Moran’s I statistic for the residuals of SAR models using different spatial weight matrices.}
\label{tab:moran_residuos}
\renewcommand{\arraystretch}{1.2}
\begin{tabular}{>{\raggedright\arraybackslash}p{6cm} 
                >{\centering\arraybackslash}p{3cm} 
                >{\centering\arraybackslash}p{3cm}}
\hline
\textbf{SAR Model} & \textbf{Moran’s I} & \textbf{p-value} \\
\hline
Queen (contiguity by vertices and edges) & 0.0151 & 0.2420 \\
Rook (contiguity by edges only) & 0.0054 & 0.3490 \\
K-Nearest Neighbors (k = 5, in UTM) & 0.0379 & 0.1410 \\
Distance Band (radius = 1000 m, in UTM) & -0.0062 & 0.4970 \\
\hline
\end{tabular}
\end{table}

\noindent These findings confirm that the SAR models were effective in capturing the underlying spatial structure of dengue incidence across neighborhoods in Recife, leaving no systematic spatial patterns in the residuals. In particular, the model using the Rook matrix stands out once again, exhibiting the lowest absolute value of spatial autocorrelation and the highest p-value—further reinforcing its robustness.

\subsection{Principal Component Analysis (PCA)}

A Principal Component Analysis (PCA) was performed using the ten selected socio-environmental variables for this study: population density, rainy season period, vegetation proportion, proportion of wealthier population, occupied households, average household size, collective housing rate, vacancy rate, proportion of canals, and proportion of informal settlements (favelas). The objective of this exploratory analysis was to assess whether dimensionality reduction through PCA could simplify the spatial modeling and mitigate potential multicollinearity issues among explanatory variables.

The PCA yielded ten principal components, which were then used as explanatory variables in a Spatial Autoregressive (SAR) model employing a Rook spatial weights matrix — previously identified as the best-performing model.

However, the model based on principal components showed a slightly lower Pseudo \(R^2\) (0.6787) compared to the model using the original variables directly (approximately 0.6864). Moreover, the interpretability of the coefficients was significantly compromised, as principal components represent linear combinations of the original variables, precluding direct association with specific socio-environmental factors.

Additionally, dimensionality reduction was not effective, since nearly all components needed to be retained to capture the relevant variability of the data, thus maintaining model complexity.

Therefore, it was concluded that the use of PCA did not provide substantial gains in predictive performance or simplification of the spatial model for dengue incidence. For this reason, the original variables were retained in the modeling to ensure greater interpretative clarity and robustness of the results.

\subsection{Spatial Modeling with SAC Models: Comparison Between Queen and Rook Matrices}

In this stage, Spatial Autoregressive Combined (SAC) models were fitted using two spatial weights matrices: Queen and Rook. The aim was to simultaneously capture spatial dependence effects in both the dependent variable and the model residuals.

Table \ref{tab:sac_metrics} presents the goodness-of-fit metrics for the SAC models. It can be observed that the SAC model with the Rook matrix exhibits slightly better performance, with a Pseudo \(R^2\) of 0.6862 compared to 0.6849 for the Queen matrix model. Additional fit criteria also favor the Rook matrix.

Table~\ref{tab:sac-coeficientes} shows the estimated coefficients for explanatory variables in both models. The signs and significance levels of coefficients are consistent between the two matrices, evidencing the robustness of the results. Noteworthy are the positive effects of population density, rainy season period, and average household size, as well as the negative effects of the proportions of canals, informal settlements, and the “wealthier” variable on dengue incidence.

Furthermore, the lambda statistic, which quantifies residual spatial autocorrelation, is lower in the model with the Rook matrix, indicating better fit and less unexplained spatial dependence. The spatial autoregressive parameter (\(W_{\text{cases\_p\_km2}}\)) was not statistically significant in either model, suggesting low direct spatial dependence of the response variable after controlling for covariates and spatial error effects.

Given these results, the SAC model with the Rook matrix is recommended for spatial analyses of dengue incidence in Recife, due to its slight statistical superiority and coherence with the territorial spatial logic of the study area. This robust modeling approach contributes to understanding socio-environmental determinants and supports the planning of public policies for disease control.

\begin{table}[H]
\centering
\caption{Goodness-of-fit metrics for SAC models with Queen and Rook matrices}
\label{tab:sac_metrics}
\begin{tabular}{@{}p{5cm}cc@{}}
\hline
\textbf{Parameter} & \centering \textbf{Queen} & \centering \textbf{Rook} \tabularnewline
\hline
Number of observations & \centering 185 & \centering 185 \tabularnewline
Number of variables & \centering 6 & \centering 6 \tabularnewline
Pseudo \(R^2\) & \centering 0.6849 & \centering \textbf{0.6862} \tabularnewline
Spatial Pseudo \(R^2\) & \centering 0.6819 & \centering \textbf{0.6831} \tabularnewline
Constant & \centering -767.03 & \centering -762.84 \tabularnewline
Lambda (spatial autocorrelation) & \centering 0.1217 & \centering \textbf{0.0482} \tabularnewline
\(W_{\text{units\_p\_km2}}\) (dependent variable autocorrelation) & \centering 0.0938 (p=0.434) & \centering 0.1338 (p=0.257) \tabularnewline
\hline
\end{tabular}
\end{table}

\begin{table}[H]
\centering
\caption{Estimated coefficients for SAC models with Queen and Rook matrices}
\label{tab:sac-coeficientes}
\begin{tabular}{@{}>{\raggedright\arraybackslash}p{3.5cm}
                >{\centering\arraybackslash}p{2cm}
                >{\centering\arraybackslash}p{2cm}
                >{\centering\arraybackslash}p{2cm}
                >{\centering\arraybackslash}p{2cm}@{}}
\hline
\textbf{Variable} & \textbf{Coef. Queen} & \textbf{p-value Queen} & \textbf{Coef. Rook} & \textbf{p-value Rook} \\
\hline
Population density       & 0.0234  & $<0.001$ & 0.0231  & $<0.001$ \\
Rainy season period      & 221.63  & $<0.001$ & 222.52  & $<0.001$ \\
Average household size   & 326.14  & $<0.001$ & 318.33  & $<0.001$ \\
Proportion of canals     & -390.01 & $<0.001$ & -372.84 & $<0.001$ \\
Proportion of informal settlements (favelas) & -228.83 & 0.0022   & -225.04 & 0.0022   \\
Wealthier population     & -184.38 & $<0.001$ & -185.92 & $<0.001$ \\
\(W_{\text{units\_p\_km2}}\) & 0.0938  & 0.434    & 0.1338  & 0.257    \\
Constant                 & -767.03 & $<0.001$ & -762.84 & $<0.001$ \\
\hline
\end{tabular}
\end{table}

\section{Discussion}

This study employed spatial econometric models—namely, the Spatial Autoregressive (SAR) and Spatial Autoregressive Combined (SAC) models—to understand and predict the geographical distribution of dengue cases in Recife over a ten-year period. Different spatial weight matrices (Queen and Rook) were tested and compared in terms of statistical performance and practical applicability.

One of the most significant findings was the high Spearman correlation ($\rho = 0.901$) between the actual ranking of accumulated dengue cases and the predicted ranking obtained from the SAR model using the Rook matrix. This result demonstrates that the model was able to accurately capture the hierarchy of the most affected neighborhoods, which is essential for territorial intervention strategies. The model correctly identified 12 out of the top 20 neighborhoods with the highest number of cases, representing a 60\% accuracy within the top 20. The SAC model exhibited similar performance, indicating that both models are comparable in terms of applicability, as confirmed by their closely aligned pseudo-$R^2$ values.

\begin{table}[H]
\centering
\caption{Comparative ranking between SAR model predictions and actual case data for the top 20 neighborhoods.}
\label{tab:ranking_sar}
\begin{tabular}{
    @{}
    >{\raggedright\arraybackslash}p{5cm}  
    >{\centering\arraybackslash}p{2cm}    
    >{\centering\arraybackslash}p{2cm}    
    >{\centering\arraybackslash}p{2.2cm}  
    @{}
}
\hline
\textbf{Neighborhood} & \textbf{SAR Rank} & \textbf{Actual Rank} & \textbf{Difference} \\
\hline
ALTO JOSE DO PINHO    & 1  & 7  & 6  \\
MORRO DA CONCEICAO    & 2  & 5  & 3   \\
AGUA FRIA             & 4  & 10 & 6  \\
MANGUEIRA             & 5  & 16 & 11 \\
BRASILIA TEIMOSA      & 6  & 18 & 12 \\
ALTO JOSE DO PINHO    & 7  & 6  & 1  \\
ALTO JOSE BONIFACIO   & 8  & 19 & 11 \\
ALTO DO MANDU         & 10 & 2  & 8  \\
BOMBA DO HEMETERIO    & 11 & 3  & 8  \\
CAMPINA DO BARRETO    & 13 & 1  & 12 \\
NOVA DESCOBERTA       & 14 & 13 & 1  \\
MORRO DA CONCEICAO    & 16 & 9  & 7  \\
TOTO                  & 21 & 4  & 17 \\
MANGUEIRA             & 27 & 11 & 16 \\
BEBERIBE              & 29 & 14 & 15 \\
COHAB                 & 32 & 20 & 12 \\
CORREGO DO JENIPAPO   & 40 & 8  & 32 \\
ALTO DO MANDU         & 45 & 15 & 30 \\
COELHOS               & 51 & 17 & 34 \\
ZUMBI                 & 56 & 12 & 44 \\
\hline
\end{tabular}
\end{table}

In addition to the rankings, thematic maps constructed from the normalized predictive scores of the SAR-Rook and SAC-Rook models displayed highly similar spatial patterns. This visual convergence reinforces the robustness of the models and confirms their usefulness as operational tools for epidemiological surveillance. These maps provide a clear visual identification of critical zones and offer a solid foundation for targeted public health actions.

It is noteworthy that the risk scores used incorporate indirect effects from neighboring regions, through the spatial term $W y$. This feature provides the models with a decisive analytical advantage: they account not only for local attributes (such as population density, precarious infrastructure, and socioeconomic characteristics) but also for spatial externalities, which better reflect the dynamics of dengue transmission in urban environments.

Although the results obtained are promising, there is room for improvement. The inclusion of new variables—such as intra-urban mobility data, road network connectivity, daily rainfall, and more granular socio-environmental indicators—could further enhance predictive accuracy. Preliminary analyses already consider the integration of seasonal climate variables, such as the effects of the El Niño phenomenon, whose influence on the incidence of arboviruses is well-documented.

From a theoretical perspective, the results highlight the importance of treating urban space as an interdependent network. The presence of spatial dependence in the data justifies the use of models that incorporate relationships between territorial units, suggesting the relevance of graph structures to represent epidemic dynamics. Based on these findings, it is feasible to extend the modeling approach to spatially coupled SIR (Susceptible-Infectious-Recovered) frameworks, where each neighborhood acts as a node with transition dynamics modulated by geographic connectivity.

Therefore, the SAR and SAC models employed not only provide reliable estimates and risk maps consistent with empirical reality, but also establish a solid foundation for integrated predictive systems and evidence-based public policy decision-making. Their applicability extends to the planning of containment strategies, resource prioritization, and scenario analysis—both in endemic contexts and during epidemic outbreaks.

\begin{figure}[H]
\centering
\begin{subfigure}{0.48\textwidth}
    \includegraphics[width=\linewidth]{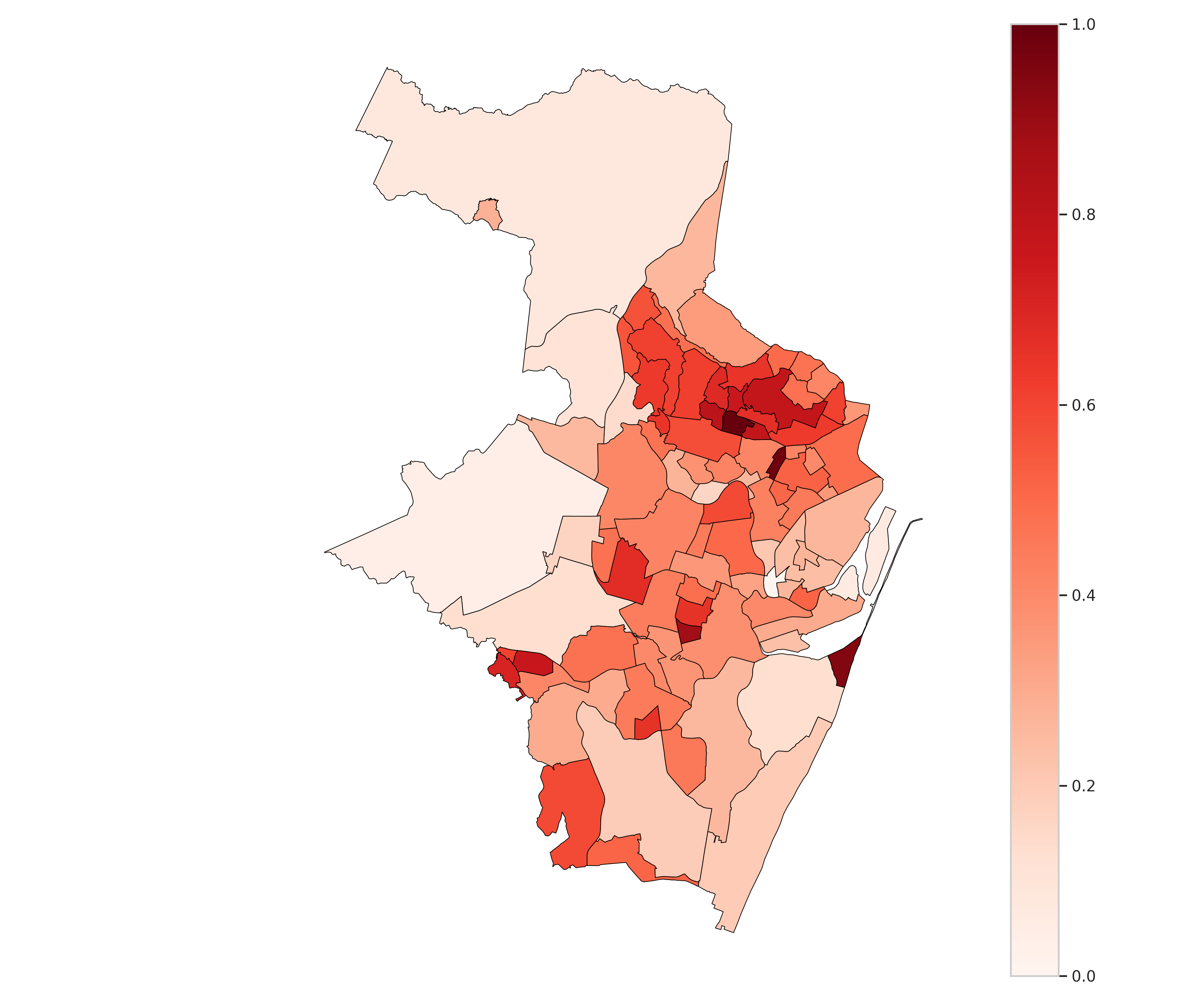}
    \caption{Thematic risk map based on the SAR-Rook model}
    \label{fig:mapa_sar_rook}
\end{subfigure}
\hfill
\begin{subfigure}{0.48\textwidth}
    \includegraphics[width=\linewidth]{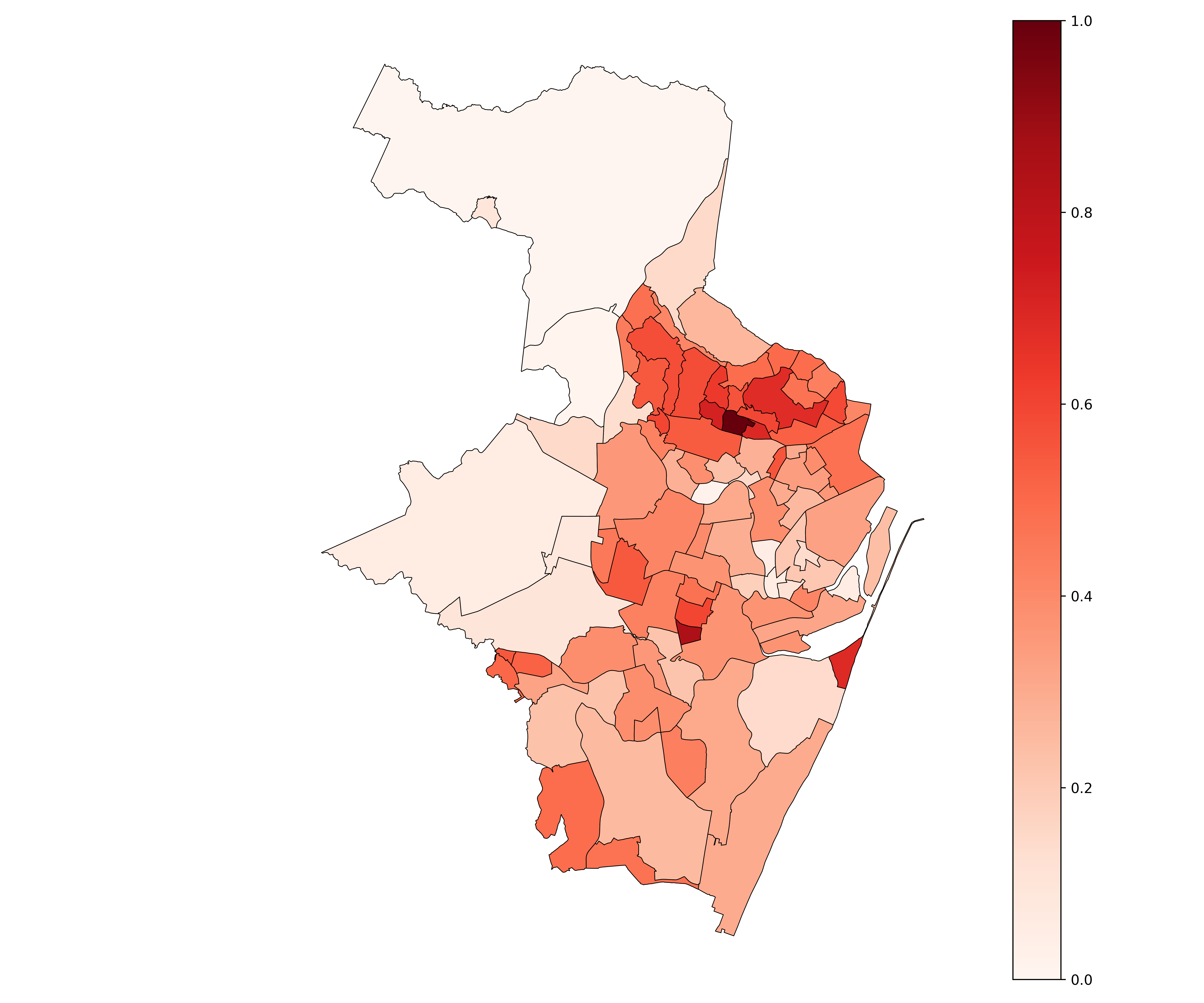}
    \caption{Thematic risk map based on the SAC-Rook model}
    \label{fig:mapa_sac_rook}
\end{subfigure}
\caption{Comparison of risk maps generated by SAR and SAC models using the Rook spatial weight matrix. Both exhibit similar spatial patterns, reinforcing model consistency.}
\label{fig:mapas_comparacao}
\end{figure}

\section{Conclusion}

This study has demonstrated that spatial econometric models such as SAR and SAC are effective tools for understanding and predicting the distribution of dengue in the urban context of Recife. The use of neighborhood matrices (Queen and Rook) revealed consistent spatial patterns of the disease, and the incorporation of spatial effects improved predictive capacity compared to traditional models.

The high correlation between predicted and observed rankings, together with the consistency of the thematic maps produced, underscores the practical utility of the proposed models. They not only provide robust estimates but can also guide public policy by prioritizing interventions in high-risk neighborhoods.

Despite these favorable results, the study acknowledges certain limitations, such as the lack of variables related to urban mobility, road network connectivity, and daily climate data. Incorporating these dimensions into future models may enhance the precision and sensitivity of spatial modeling.

In summary, the SAR and SAC models represent a solid foundation for integrating spatial intelligence into epidemiological surveillance and territorial health management. They are applicable to other urban settings and vector-borne diseases, paving the way for the development of early warning systems grounded in geospatial evidence.

\end{document}